\documentclass{article}
\usepackage{spconf,amsmath,graphicx}
\usepackage{graphicx}
\usepackage{caption}
\usepackage{subcaption}

\usepackage{enumitem}
\setlist{nosep, leftmargin=14pt}

\usepackage{mwe} % to get dummy images

\title{$MHATC$: Autism Spectrum Disorder identification  utilizing multi-head attention encoder along with temporal consolidation modules}
\name{Ranjeet Ranjan Jha$^{*1}$, Abhishek Bhardwaj$^{1}$, Devin Garg$^{2}$, Arnav Bhavsar$^{1}$, Aditya Nigam$^{1}$%\thanks{Some author footnote.}
}
\address{$^{1}$Indian Institute of Technology Mandi, $^{2}$Indian Institute of Technology Jodhpur}
\begin{document}
%\ninept
%
\maketitle
\begin{abstract}
Resting-state fMRI is commonly used for diagnosing Autism Spectrum Disorder (ASD) by using network-based functional connectivity. It has been shown that ASD is associated with brain regions and their inter-connections. However, discriminating based on connectivity patterns among imaging data of the control population and that of ASD patients' brains is a non-trivial task. In order to tackle said classification task, we propose a novel deep learning architecture (MHATC) consisting of multi-head attention and temporal consolidation modules for classifying an individual as a patient of ASD. The devised architecture results from an in-depth analysis of the limitations of current deep neural network solutions for similar applications. Our approach is not only robust but computationally efficient, which can allow its adoption in a variety of other research and clinical settings.
\end{abstract}

\section{Introduction}
Autism spectrum disorder (ASD) is considered as one of the most complex neuro-developmental disorders, in which patients suffer from several difficulties, including repetitive behavior, deficits in social communication, and restricted interest. Over the years, functional Magnetic Resonance Imaging (fMRI) has been established as a well-known sensing technique and proved to be useful in assessing different mental disorders. Utilizing fMRI, it has been observed that the functional connectivity (FC) among the different regions of the brain can yield discriminative representation between control and a patient of ASD. In the literature, several works have been reported, which primarily operate on the FC matrix.
%This connectivity is typically computed by obtaining and further processing the temporal correlation between activities of all pairs of regions in the brain. In the literature, several works have been reported, which primarily operate on this functional connectivity matrix. 
%among other techniques such as Electroencephalography (EEG), Magnetoencephalography (MEG), and Positron Emission Tomography (PET), and has been proved to be useful in assessing different mental disorders. 

Early detection of ASD is very crucial for better treatment of patients. However, this is a very tedious and time-consuming task for medical practitioners as several clinical observations need to be made about a patient's behaviour to diagnose this mental disorder. In this regard, fMRI, along with sophisticated signal processing techniques, can aid in the early identification of ASD. In addition, several machine learning-based techniques have been reported in the literature to perform the underlying classification task (Control vs ASD); but, those techniques depend on hand-crafted feature extraction methods, which limit the performance. 
%The use of a hand-engineered feature extraction procedure seems to be an evident reason behind the limited performance of the ML techniques mentioned above.
%Several machine learning-based techniques have been reported in the literature to perform the underlying classification task (Control vs Autism); but, those techniques depend on hand-crafted feature extraction methods. Most of the earlier works have utilized the correlation matrix to extract features for machine learning (ML) classifiers, including random forest, naive Bayes, and support vector machine (SVM) approaches \cite{retico2016effect}. The use of a hand-engineered feature extraction procedure seems to be an evident reason behind the limited performance of these ML techniques mentioned above.

Recently, encouraged by the popularity of deep learning methods in the computer vision domain and sub-domains, few researchers have deployed these techniques also, in classification tasks involving the fMRI data including that of ASD identification. As deep learning techniques are end to end, we do not perform the feature extraction separately, unlike traditional machine learning, where this is the backbone. In \cite{2}, the authors have utilized a Siamese graph convolutional neural network (GCNN) to perform the ASD classification (as opposed to the vanilla 3D-CNN approach taken by Khosla \textit{et al.} in \cite{khosla20183d}). They worked with graph convolution in the spectral domain and achieved comparable results to state-of-the-art techniques. In \cite{parisot2018disease}, authors have considered a GCNN for the identification of ASD and Alzheimer's disease. Heinsfeld \textit{et al.} \cite{heinsfeld2018identification} have used an encoder-decoder based architecture for getting robust and discriminative low-dimensional features from the FC matrix, which consequently help in better classification. Taking a temporal approach, Dvornek \textit{et al.} \cite{dvornek2017identifying} have performed the classification of individuals with ASD and typical controls by directly employing LSTM modules on resting-state fMRI time series data. This led to a significant increase of 9\% in the classification accuracy at the time, as reported by them.

\begin{figure*}[!ht]
     \centerline{\includegraphics[width=1.0\linewidth, height=2.9cm]{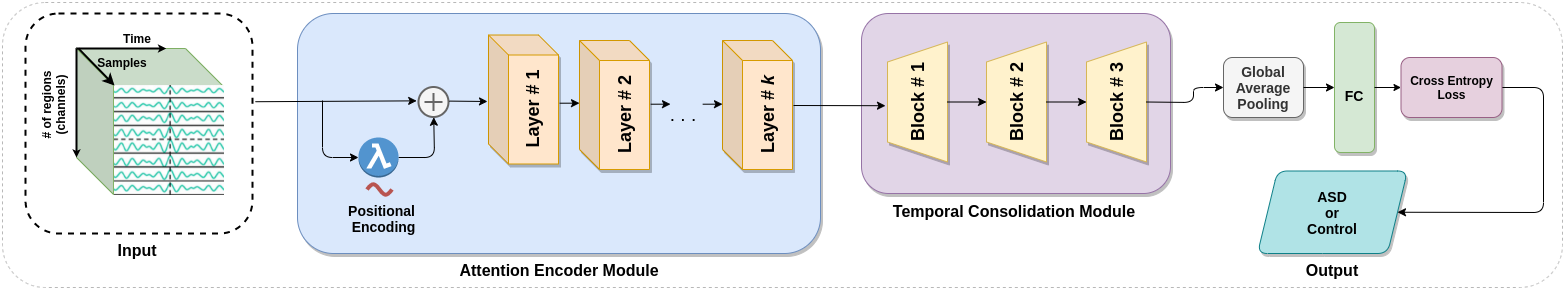}}
     \centering
       \caption{Proposed Architecture}
        \label{fig:isbi2}
        \vspace{-0.4cm}
\end{figure*}
Authors in \cite{zhao2018diagnosis} have employed multi-level high order functional connectivity, to get a more comprehensive characterization of the complex interactions between the different regions of the brain. Feature selection to identify discriminative FC features in their approach led to improved ASD identification results. In \cite{niu2020multichannel}, an attention-based deep learning model has been proposed which achieved good accuracy. In \cite{jiao2020improving} and \cite{huang2020identifying}, a deep learning-based capsule network and a Deep Belief Network, respectively, have been presented for performing the ASD classification task. Even the combination of sMRI and fMRI has been explored for autism spectrum disorder diagnosis in \cite{7950683} where the decisions made by both the modalities have been fused using a deep autoencoder network. In \cite{ingalhalikar2021functional},ASD prediction has been performed on site harmonized autism dataset. \cite{jung2021inter} have tried to learn inter regional high level relation from functional connectivity for the classification of ASD disease.

However, the problem is still challenging and even  after many such works there is still considerable scope to improve the overall classification accuracy. Most of the state-of-the-art deep learning techniques, proposed for ASD classification, directly using the FC matrix as input which approximates the data at an initial stage with second-order statistics. On the other hand, our model works on the original input signal, and learns robust features by employing a multi-head self attention and temporal consolidation mechanism.

 %Since, wrongly estimated hand-crafted features may mislead the deep learning based classification techniques, our model focuses on learning robust and useful features from the functional connectivity matrix implicitly, by employing a multi-head self attention mechanism.

%The overall pipeline of autism classification is given in Fig. \ref{fig:isbi1}. Using a well established brain parcellation (into $200$ regions) \cite{craddock2012whole}, the average rs-fMRI signal for each region is computed, where each signal has $200$ time points. Thus, we have obtained the matrix $S$ of size $200$ (\# regions) $\times$ $200$ (\# time points). The proposed technique uses this matrix for performing the classification task between autism and control class.  

%We have proposed the novel architecture, which works on input $ S $ and performs the classification task between autism and control.
Our contribution in this paper is of several folds:

(1) Classification of ASD utilizing a novel $MHATC$ architecture (as shown in Fig. \ref{fig:isbi2}), consisting of a  multi-head attention encoder ($MHAE$) and a temporal consolidation module (TCM). (2) The $MHAE$ (Fig. \ref{fig:isbi4}) which includes a positional encoding component and  leverages several blocks of multi-head attention and feed-forward networks with residual connections for flexibility. (3) The TCM (Fig. \ref{fig:tcm}) which includes several context consolidation modules that aggregate the features across regions at any given time step, which results in a better overall feature representation. (4) Achieving state-of-the-art performance on the publicly available ABIDE data-set.

\section{Method}
We have proposed a cascaded network $MHATC$ (as shown in Fig. \ref{fig:isbi2}), which consists of two major modules - the Multi-Head Attention Encoder (MHAE) module and the Temporal Consolidation Module (TCM). As the name suggests, the MHAE module is employed to perform attention based encoding of the features required for classification (inspired by \cite{vaswani2017attention}). On the other hand, the TCM is utilised to step-by-step consolidate the concerned features across the time points and get a condensed feature map at the end. Finally, a global pooling layer, followed by the classification layer, provides the final classification output. 

%Thus, the proposed network's overall goal is to learn temporal information about the functioning of the brain regions individually as well as the relationship among them.

\subsection{Multi-Head Attention Encoding Module} 
In general, a $MHAE$ learns position invariant representation of the input data as discussed in \cite{vaswani2017attention}. However, for temporal data, positional information within the given input sequential data is important. This gives rise to the need of positional information to be embedded in the input. For this purpose, we have employed sinusoids like in \cite{vaswani2017attention} to associate positional information with the input data, where phase of the sinusoids varies for different dimensions of the encoding. The input to this module is the matrix of size (200,200) where we have data for 200 time-points for every region and the brain is divided into 200 regions. The module functions by first calculating the positional embedding of the input block which helps attach a temporal meaning to the input fMRI data. Post this, the embedding is added to the input and passed to an encoder which employs multi-head self attention followed by a feed-forward network to capture the relative importance of features in the input. Multiple such layers, each including multi-head attention and feed-forward components, have been used which has shown to improve classification performance. Note that there is no reduction in size as the input passes through this module. 

% \textbf{Positional Embedding:} Positional embedding tends to incorporate ordering and provides a unique encoding for each time-step. For attaching a temporal meaning to the input by means of this positional embedding, the sinusoidal function has been utilized, where $sin(wt)$ and $cos(wt)$ have been used for even and odd places of the input vector's dimensions, respectively. 
% %In this way, the positional encoding step transforms one temporal sequence into another sequence.

% \begin{equation}
%     \label{sin}
%     P(pos, 2i) = sin(\frac{pos}{10000^{2i/k}})
% \end{equation}
% \begin{equation}
%     \label{cos}
%     P(pos, 2i+1) = cos(\frac{pos}{10000^(2i/k)})    
% \end{equation}

% In the equations \ref{sin} \& \ref{cos}, pos denotes the position, i is the dimension whose encoding is being calculated and k is the dimension of the input to the multi-head attention unit. Note that every dimension of such an encoding is representative of a sinusoid.
\begin{figure*}[!ht]
    \centering
    \begin{subfigure}{0.25\textwidth}
        \centering
        \includegraphics[width=\textwidth, height=5.5cm]{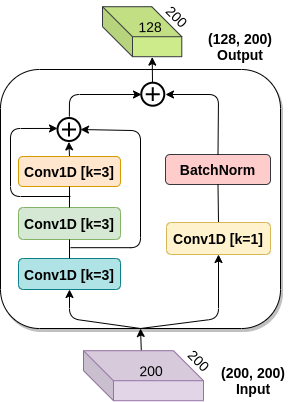}
        \caption{A block in the Temporal Consolidation Module. }
        \label{fig:tcm}
    \end{subfigure}
    \hfill
    \begin{subfigure}{0.74\textwidth}
        \centering
        \includegraphics[width=\textwidth, height=5.5cm]{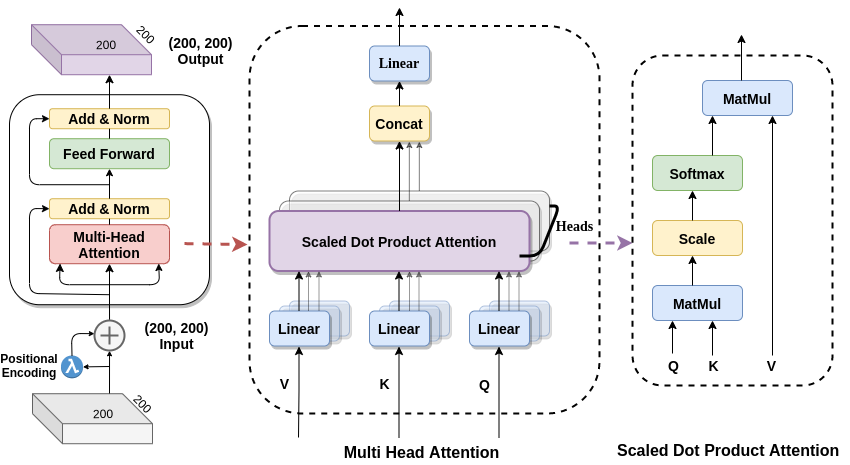}
        \caption{A layer in the Multi-Head Attention Encoder module. Multiple such layers have been employed in a cascaded manner for improved encoding of inter-region dependence.}
        \label{fig:isbi4}
    \end{subfigure}
    \caption{Note that in (a), a TCM block causes a (200,200) input to be consolidated to a (128,200) output owing to 1D convolutions with filter size ($k$) of 3. Multiple such blocks employed in a cascaded manner cause the consolidation to become further dense with the size reducing to (64,200) and then finally to (32,200). In (b), the MHAE module is shown.}
    \vspace{-0.4cm}
\end{figure*}

\textbf{Self-attention: }This module's core idea lies in \textit{multi-head self-attention}, which corresponds to the ability to attend to different positions of the input sequence to compute a representation of that sequence. We have used a stack of multi-head self-attention layers along with feed-forward neural networks. For a self-attention layer, there are three input vectors, viz. query ($q$), key ($k$), and value ($v$) \cite{vaswani2017attention} which indicate the role that each element of the input sequence plays. Here, three representations $q$, $k$ and $v$ have been obtained from transformation of input $I$ (after positional embedding).
The degree of correlation between two elements in a piece of sequential data determines the extent to which these elements influence the output at a given time-step. The attention ($A$) (similar to \cite{vaswani2017attention}) is given by the following relation:
\begin{equation}
    \label{attention_1}
    A' = \frac{q \; \cdot k}{\sqrt{m}}
\end{equation}
\begin{equation}
    \label{attention_2}
    A = softmax(A')
\end{equation}
\begin{equation}
    \label{attention}
    F = A \: \cdot v = softmax(\frac{q \: k^{T}}{\sqrt{m}})\; \cdot v
\end{equation}
%Since dot-product attention is time-space efficient, we have considered this attention mechanism for our experiments. 
In our case, the query and key have the same dimension, i.e., $m$ which facilitates the dot product in the above equations. This dot-product is followed by scaling i.e. division by square root of m, resulting in an $m$-dimensional vector $A'$ (eq. \ref{attention_1}). This has a normalizing effect and prevents explosion of values. Subsequently, softmax function has been applied on $A'$ to get the attention vector as the upshot i.e. $A$ which holds different weightage for different entries in the value vector. Finally, this attention $A$ is multiplied with the value vector $v$ to get the feature vector $F$. 

\textbf{Multi-head attention:} To jointly attend to information at different positions from different representational spaces, multi-head attention has been employed. Multi-head attention, to put it simply, involves the  concatenation of outputs by multiple simple scaled dot-product attention units as shown in Fig. \ref{fig:isbi4}. This concatenated output is finally passed through a dense layer. We have used five such `layers' in the module, where the multi-head attention component of each layer consists of two attention heads. 
%A point to note here is that the attention scheme employed here offers a great degree of parallelization which leads to a more efficient handling of an otherwise sequential-natured data. %Multi-head attention i.e. linearly projecting a (q,k,v) head to multiple dimensions provides the ability for the model to itself decide which combinations of attention to focus on, since these projections are weighted with learnable parameters too. The effects of a change in the number of layers in this cascaded structure have been discussed in the ablation study.
\subsection{Temporal Consolidation Module}
%The blocks in this module are inspired by the modules used in the Inception Network \cite{szegedy2015going}.
This module aims to extract the more discriminative features by aggregating the brain's regions and capturing the multi-context information. One can observe in Fig. \ref{fig:tcm} that we have concatenated three types of feature maps taken at three contexts by applying filters of size $3$, $5$ (two sequential $3$), and $7$ (three sequential $3$), respectively. This module's input dimension has been reduced to $128 \times 200$ by including $128$ filters in the first 1D convolution layer; however, the number of filters remains the same for other layers. In this way, the final output after the multi-context layer would be $M_1$ of size  $128 \times 200$, which contains the inherent multi-context representation. Further, there is an additional residual connection branch that contains only one 1D convolution layer of filter size $1$. This layer's task is to forward the input directly to the multi-context output; nevertheless, the dimension would be changed to $128 \times 200$ for addition with $M_1$. This technique is used as it is known to speed up gradient flow during backpropagation. An important thing that has been exploited here is the ability of the network to decide what layers it requires for a well-enough representation of the crucial features. This adds to the flexibility of the network in not being rigid about the types of filters to be used, but allows for automatic selection due to the end-to-end nature. Subsequently, two more TCM blocks have been included in sequential order with the same block structures and the different number of filters (64 and 32). Thus, the aggregated feature map only contains the most crucial relationship information among the different regions, and hence helps in better classification.

\section{EXPERIMENTAL SETUP}
\subsection{Dataset description}
For the evaluation of our proposed network, the complete ABIDE I \cite{nielsen2013multisite} fMRI dataset has been utilized. This dataset has contributions from multiple different international brain imaging laboratories. It consists of 530 control (normal) individuals and 505 ASD patients. We have performed 10-fold cross validation like in \cite{huang2020identifying}, for checking the robustness of the model. %This whole dataset has been divided into training, validation and testing sets which are 80, 10 and 10 percent of the whole, respectively.
\subsection{Evaluation metrics and implementation details}
%The accuracy metric has been used for which evaluations have been discussed in this section. Accuracy is a measure of the predictive ability of the model. Sensitivity is just another term for the `hit rate' which indicates how correctly the model predicts a person with autism. Specificity, on the other hand, denotes how well the model classifies individuals into the other class i.e. typical controls. In addition, ROC curve has been plotted, where the area under the curve (AUC) is another important measure. ROC curves provide a good visual measure of identifying the discriminative ability of a model. 
The accuracy, sensitivity and specificity metrics similar to \cite{jiao2020improving}, have been used for which evaluations have been discussed in this section. In addition, ROC curve has been plotted, where the area under the curve (AUC) is another important measure. 
%ROC curves provide a good visual measure of identifying the discriminative ability of a model.

%Mathematically, the accuracy is defined as follows:

%\begin{equation}
%    \label{accuracy}
%    Accuracy = \frac{\#\:Correctly\:\:classified\:\:samples}{\#\:Total\:\:samples}
%\end{equation}

 %\begin{equation}
 %    \label{recall}
 %    Sensitivity = \frac{\#\:True\:Positives}{Actual\:\#\:Positives\:in\:the\:population}
 %\end{equation}

 %\begin{equation}
 %    \label{spec}
 %    Specificity = \frac{\#\:True\:Negatives}{Actual\:\#\:Negatives\:in\:the\:population}
 %\end{equation}

\section{Results and Discussion}
\begin{figure}[ht]
%\begin{subfigure}{0.5\textwidth}
%  \centering
  % include second image
%  \includegraphics[width=1.0\linewidth]{epoch_validation_plot.png}  
%  \caption{Epoch versus Validation Accuracy}
%  \label{fig:val}
%\end{subfigure}
%\begin{subfigure}{0.5\textwidth}
 \centering
  % include first image
  \includegraphics[width=0.9\linewidth, height=5.6cm]{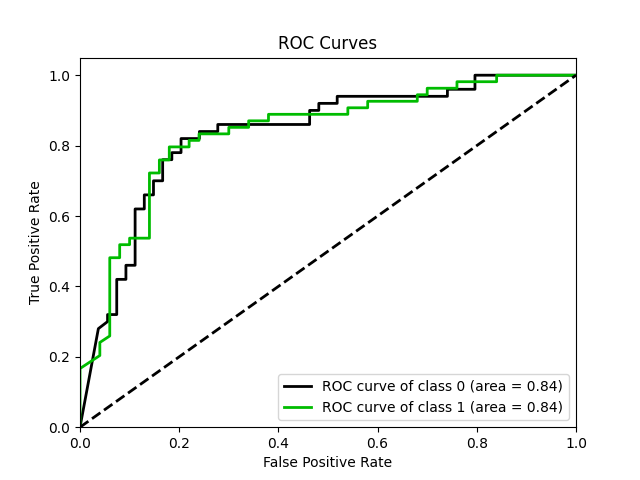}  
\caption{Receiver Operating Characteristic (ROC) curves (class 0 for control and class 1 for the patient)}
\label{fig:rv}
\vspace{-0.7cm}
\end{figure}

% \begin{table}[!htp]
% \begin{center}
% \caption{Metrics measured for different testing databases}
% \label{tab:resultper}
% \begin{tabular}{|c|c|c|c|c|}
% \hline
%   \textbf{Methods} & \textbf{Accuracy} & \textbf{Sensitivity} & \textbf{Specificity} \\ \hline 
% \textbf{Proposed method} & 80.56\% & 65.11\% & 90.77\%  \\ \hline
% \end{tabular}
% \end{center}
% \end{table}
%In Fig. \ref{fig:val}, validation accuracy plot has been depicted, where one could see that the best performance has been achieved before $200$ epochs only. 
We provide a comparison of the accuracy of our approach with the present state-of-the-art methods. 
%Note that the accuracy is considerably improved (at 77.4\%) compared to previous best performing methods.
Table \ref{tab:compare} provides the average performance analysis for cross validation. It can be observed that our proposed model is giving good mean accuracy, i.e., $77.4\%$. Another point to note is that, mean sensitivity is high ($78.6\%$), and such a high sensitivity enables correct identification of patient individuals to a great extent. However, mean specificity is comparatively less ($74\%)$. The reason behind this can be that rs-fMRI signals for ASD individuals have more discriminative representation than that of controls. In addition, we have shown the best ROC curve among all the validation data (considered for each fold) for ASD and control classification in Fig. \ref{fig:rv}, and the mentioned value of AUC also indicates a high-quality performance of our method.
%we  have proposed
% Note that the specificity value is a measure of the probability of correct classification of a control individual given the individual's rs-fMRI data. This is a desirable measure for any kind of diagnosis system as we tend to try and reduce the number of false positives (as a higher specificity value would lead to lesser false positives). These measures are graphically represented using the ROC curve.
In addition to the overall accuracy, an ROC curve helps in providing more insight into the performance of the model. It helps figure out the trade-off between false positives and false negatives according to the use-case of the system under scrutiny. 

%\begin{figure}[!h]
 %    \centerline{\includegraphics[width=0.5\linewidth, height=4cm]{roc2.png}}
  %   \centering
   %    \caption{}
    %    \label{fig:roc}
%\end{figure}
\begin{table}[!ht]\scriptsize
 \caption{Comparative Analysis and Ablation Study}
    \begin{subtable}[h]{0.30\textwidth}
        \begin{tabular}{|c|c|c|c|}
\hline
  \textbf{Model} & \textbf{Accuracy}  & \textbf{Sensitivity} & \textbf{Specificity} \\ \hline 
    Ghiassian \textit{et al.} \cite{ghiassian2016using} & 59.2\% & 72.2\% & 45.4\% \\ \hline
    %Nielsen \textit{et al.} \cite{nielsen2013multisite} & 60.0\% & 62.0\% & 58.0\%\\ \hline
    Chen \textit{et al.} \cite{chen2015diagnostic} & 66.0\% & 60\% & 72\% \\ \hline
   % Abraham \textit{et al.} %\cite{abraham2017deriving} & 66.9\% & - & - \\ \hline
   % Dvornek \textit{et al.} \cite{dvornek2017identifying} & 68.5\% & - & - \\ \hline
%    Plitt \textit{et al.} \cite{plitt2015functional} & 69.7\% & - & \\ \hline
    Heinsfeld \textit{et al.} \cite{heinsfeld2018identification} & 70.0\% & 74\% & 63\%\\ \hline 
    
   Jiao \textit{et al.} \cite{jiao2020improving} & 71\% & 73\% & 66\% \\ \hline
    Huang \textit{et al.} \cite{huang2020identifying} & 76.4\% & 77.8\% & 75.0\% \\ \hline 
    
    Ingalhalikar \textit{et al.} \cite{ingalhalikar2021functional} & 71.35\% & 59.5\% & 80.6\% \\ \hline 
     Jung1 \textit{et al.} \cite{jung2021inter} & 69.19\% & 64.79\% & 73.46\% \\ \hline 
    \textbf{Proposed method} &  \textbf{77.4}\% & \textbf{78.6}\% & \textbf{74}\% \\ \hline 
    
\end{tabular}
       \caption{Performance comparison between our proposed network and state-of-the-art-works}
       \label{tab:compare}
    \end{subtable}
    \hfill
    \begin{subtable}[h]{0.55\textwidth}
            \begin{tabular}{|c|c|c|}
\hline
  \textbf{\# Encoders} &  \textbf{\# Attention heads} & \textbf{Accuracy}  \\ \hline 
     1 & 2 &  73.56\% \\ \hline
     2 & 2 &  75.48\% \\ \hline
     3 & 2 &  76.62\% \\ \hline
     4 & 2 &  77.00\% \\ \hline
      \textbf{5} &  \textbf{2} &  \textbf{77.40}\%  \\ \hline
     6 & 2 & 76.80\% \\ \hline
    % 7 & 4 & 76.82\%\\ \hline
    
\end{tabular}
        \caption{Ablation Study: Comparison for multi-head attention encoder \\ having different combinations of attention heads and encoder layers}
        \label{ablation}
     \end{subtable}
     \label{tab:temps}
     \vspace{-0.5cm}
\end{table}

%\textbf{Comparative Analysis}
% For comparative analysis, we have considered a similar strategy as taken by all the previously existing methods. 
% The performance comparison has been shown in Table \ref{tab:compare}, where our method is providing the best validation accuracy, i.e., $80.56\%$. In this way, we can ensure that our approach is more generalizable towards the classification between control and patients.

%\begin{table}[!htp]
%\begin{center}
%\caption{Performance comparison between our proposed network and state-of-the-art-works}
%\label{tab:compare}
%\begin{tabular}{|c|c|}
%\hline
%  \textbf{Model} & \textbf{Accuracy}  \\ \hline 
 %   Ghiassian \textit{et al.} \cite{ghiassian2016using} & 59.2\% \\ \hline
  %  Nielsen \textit{et al.} \cite{nielsen2013multisite} & 60.0\% \\ \hline
   % Chen \textit{et al.} \cite{chen2015diagnostic} & 66.0\% \\ \hline
%    Abraham \textit{et al.} \cite{abraham2017deriving} & 66.9\% \\ \hline
%    Dvornek \textit{et al.} \cite{dvornek2017identifying} & 68.5\% \\ \hline
%    Plitt \textit{et al.} \cite{plitt2015functional} & 69.7\% \\ \hline
%    Heinsfeld \textit{et al.} \cite{heinsfeld2018identification} & 70.0\% \\ \hline 
%    Huang \textit{et al.} \cite{huang2020identifying} & 76.4\% \\ \hline 
%    \textbf{Proposed method} & 79.63\% \\ \hline 
    
%\end{tabular}
%\end{center}
%\end{table}

\begin{figure}[!ht]
     \includegraphics[width=0.9\linewidth, height=6.5cm]{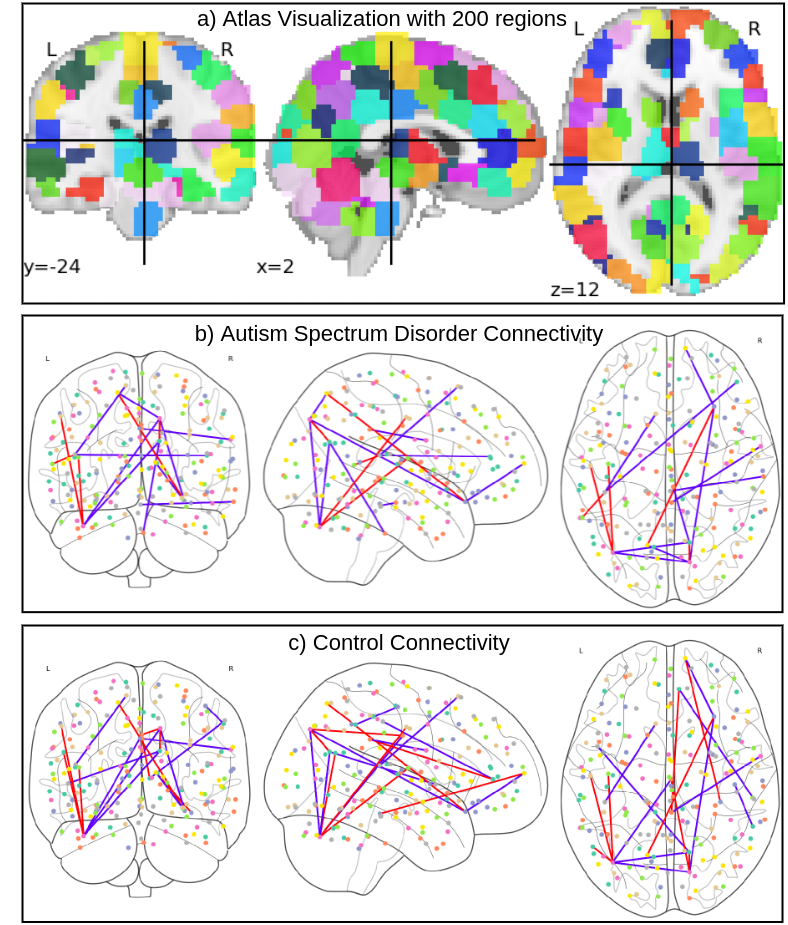}
     \centering
       \caption{Atlas Visualization with 200 regions and Selected few highly correlated edges in both the ASD and Control (in coronal, sagittal and axial view)}
        \label{fig:atlas}
        \vspace{-0.3cm}
\end{figure}

% \textbf{Connectivity Visualization:}
For a visual assessment, we have shown the Craddock atlas \cite{craddock2012whole} in Fig. \ref{fig:atlas}, where there are $200$ regions with color encoding for coronal, sagittal and axial view. As we know, that the $MHAE$ module has been used to get refined features for each region; we have considered the mean features of the $MHAE$ module for ASD and control classes separately. Further, including only the edges which have value higher than $0.6$ (this value being an indicator of the extent to which two regions in the brain are connected functionally) have been plotted in Fig. \ref{fig:atlas}. One can observe that there is a difference in the regions which are `connected' i.e. are engaged in some functionality together in case of ASD and control. This shows the discriminability of the $MHAE$ features. The result depicts the involvement of several regions, including left fusiform gyrus, right precentral gyrus, left superior frontal
gyrus, left precuneus cortex, for distinguishing ASD from control.
Thus, we can say that the $MHAE$ module aids in extracting discriminative features. 

%\begin{figure}[ht]
%\begin{subfigure}{0.5\textwidth}
%  \centering
  % include second image
%  \includegraphics[width=1.0\linewidth, height=4.0cm]{Autism.png}  
%  \caption{Autism Connectivity}
%  \label{fig:val}
%\end{subfigure}
%\begin{subfigure}{0.5\textwidth}
% \centering
  % include first image
%  \includegraphics[width=1.0\linewidth, height=4.0cm]{Normal.png}  
%  \caption{Control Connectivity}
%  \label{fig:roc}
%\end{subfigure}
%\caption{Selected few highly correlated edges in both the Autism and Control}
%\label{fig:connectivity}
%\end{figure}

%\textbf{Ablation Study: }
%Considering the Multi-Head Attention Encoding module, we have analyzed the effect of varying the number of encoders as well as attention heads on the classification between control and autism. In Table \ref{ablation}, it can be noticed that varying number of attention heads upto two while fixing \# encoder layers increases the accuracy. Thus, multi-head attention helps a lot in raising overall performance. However, if we increase more than two heads, there is no improvement in accuracy. Moreover, going with more than four encoders does not help in boosting the performance. Hence, four encoders each containing two attention heads have given the best performance among all the combinations. 

Considering the Multi-Head Attention Encoding module, we have analyzed the effect of varying the number of encoders on the classification between control and ASD. During the experiment, we have observed that if we increase more than two attention heads, there is no improvement in accuracy. Moreover, going with more than five encoders reducing the performance. Hence, five encoders, each containing two attention heads, have given the best performance among all the combinations. 
\section{Conclusion}
We have proposed an end to end deep learning architecture that leverages several modules, including multi-head attention and temporal consolidation, to perform the classification task between control and autism spectrum disorder fMRI scans. Our approach yields state-of-the-art accuracy and is also quite robust and computationally efficient. Thus, one can explore its application to various other research and clinical settings. In the future, we would explore variations of the proposed network to improve the performance further.

\section{Compliance with Ethical Standards}
The above work has been conducted using human subject data made available in open access by the Autism Brain Imaging Data Exchange. Ethical approval was not required as confirmed by the license attached with the open access data.

\bibliographystyle{IEEEtran}
\bibliography{strings}

\end{document}